\documentclass[12pt,a4paper]{article}

\usepackage{amsmath}
\usepackage{amssymb}
\usepackage{tikz}
\usepackage{fixltx2e}
\usepackage{amsfonts}
\usepackage{graphicx}
\usepackage{subfig}
\usepackage{caption}
\usepackage{dcolumn}
\usepackage{bm}
\usepackage[T1]{fontenc}
\usepackage{textcomp}
\usepackage[
colorlinks,linkcolor=blue,citecolor=magenta]{hyperref}
\usepackage[top=3cm,right=2cm,bottom=2.5cm,left=2cm]{geometry}
\begin{document}
\title{Structure of Constrained Systems in Lagrangian Formalism and Degree of Freedom Count}

\author{Mohammad Javad Heidari${}^{a}$\thanks{Jheidari840@gmail.com}, Ahmad Shirzad${}^{a,b}$\thanks{shirzad@ipm.ir}\\
${}^a$ \textit{Department of Physics, Isfahan University of Technology}
\\${}^b$ \textit{School of Particles and Accelerators,}
\\\textit{Institute for Research in Fundamental Sciences
(IPM)}, \\\textit{P.O.Box 19395-5531, Tehran, Iran}}
\maketitle
\begin{abstract}
A detailed program is proposed in the Lagrangian formalism to investigate the dynamical behavior of a theory with singular Lagrangian. This program goes on, at different levels, parallel to the Hamiltonian analysis. In particular, we introduce the notions of first class and second class Lagrangian constraints. We show each sequence of first class constraints leads to a Neother identity and consequently to a gauge transformation. We give a general formula for counting the dynamical variables in Lagrangian formalism. As the main advantage of Lagrangian approach, we show the whole procedure can also be performed covariantly. Several examples are given to make our Lagrangian approach clear.	
\end{abstract}

\section{Introduction}
Since the pioneer work of Dirac \cite{dirac} and subsequent forerunner papers (see Refs. \cite{sundermeyer,henn,Batlle} for a comprehensive review),  people are mostly familiar with the constrained systems in the framework of Hamiltonian formulation. The powerful tool in this framework is the algebra of Poisson brackets of the constraints. As is well-known, the first class constraints, which have weakly vanishing Poisson brackets with all constraints, generate gauge transformations. However, there is no direct relation, in the general case, to show how they do this job. In other words, a complicated procedure is required to construct the generator of gauge transformation (see chapter 3 of ref. \cite{henn}) by   suitably combining the first class Hamiltonian constraints and arbitrary functions of time (or space-time for field theories). On the other hand, the price to be paid for using advantages of the Hamiltonian formulation is breaking the manifest covariance of the system. 

The famous formula for the number of dynamical degrees of freedom in phase space is \cite{gomhenpons}
\begin{equation}
D^H=(2K)- 2F^H-S^H\label{01}
\end{equation}
where $(2K)$ is the dimension of the phase space and $F^H$ and $S^H$ denote the number of  first class and second class Hamiltonian constraints respectively. Remember that the second class constraints are those with a nonsingular matrix of mutual Poisson brackets among each other. 

 In ref. \cite{Batlle} it is shown that projecting primary Hamiltonian constraints into Lagrangian variables gives identically zero, while this procedure leads to Lagrangian constraints for subsequent levels of Hamiltonian constraints. Some aspects of the  constrained systems in the Lagrangian formalism are also studied in refs. \cite{gitman,chaichian}. It is also well-known that the null-vectors of the Hessian matrix lead to primary Lagrangian constraints \cite{sudarshan}.

In ref. \cite{dr} a certain method is introduced to follow the consistency procedure of the Lagrangian constraints. This method is based on constructing the extended Hessian by adding newer lines at the bottom of the Hessian matrix which correspond to time derivatives of the Lagrangian constraints. Moreover, it is shown that if a new null-vector of the extended Hessian does not lead to a new constraint, it would lead to a Neother identity. Finally, it is shown that every Neother identity may be written in such a form which enables us to recognize directly the so called gauge generators. 

The methods given in ref. \cite{dr} is used more or less when people are interested in Lagrangian investigation of a gauge system. For example, in ref. \cite{banerjee} the Hamiltonian gauge generators are compared with their counterparts in a purely Lagrangian approach. In ref. \cite{ba roy sa} the Poincare gauge theory formulation of gravity is studied in the context of the purely Lagrangian approach. This approach is also employed in ref. \cite{bansam} in studying the gauge transformations and the corresponding generators on a non-commutative space.

In this paper we want to give a complete Lagrangian program for investigation the physical properties of a constrained system. We show  (for the first time) that we can classify the Lagrangian constraints into first class and second class ones. We show that the Lagrangian constraints may be managed as constraint chains, analogues to the Hamiltonian constraint chains introduced in \cite{loranshirzad}. Each first class chain leads finally to a Neother identity which introduces one gauge parameter (and its time derivatives) in the solutions of equations of motion. For field theories each Neother identity introduces an arbitrary field and its space-time derivatives in the solutions. Similar to the formula (\ref{01}) we will derive an important formula to find the number of dynamical degrees of freedom in Lagrangian formalism (see Eq. \ref{No.dyn} below). 

The main structure of our Lagrangian approach is given in the 
following two sections which also include the main features of the method given in ref. \cite{dr}. This structure is explained for systems with finite number of degrees of freedom. In section 2 we classify constraints at each level of consistency into three different classes, i.e. first class, second class and pending constraints. We denote this procedure as "FPS decomposition". This structure finally deposits a number of first class and a number of second class Lagrangian constraint chains with different lengths. In section 4 we give a few examples to make the whole approach more comprehensible. The first three examples are simple toy examples which may help the reader to get the ideas of the sections 2 and 3 more rapidly. 

For field theories, however, we may have two approaches. In the first approach, we may depart from manifest covariance of the theory and treat the time as the distinguished evaluation parameter of the dynamical system. Hence, the space coordinates act as continues labeling parameters of the dynamical variables. This approach, as is well-known, is parallel to Hamiltonian investigation which breaks the apparent covariance of the system. In the last part of section 4 we investigate the electromagnetic theory in non-covariant approach. This method however may lead to lengthy and tedious calculations involving so many components of the tensor fields. For instance, working out the Einestain- Hilbert action in this non covariant method includes so many pages of ref. \cite{samanta}. 

In Lagrangian formalism, a second possibility for field theories, is the covariant treatment of the dynamical equations of motion. For regular field theories (without constraints), such as scalar field theory, the ordinary Euler-Lagrange equation is replaced obviously by a covariant equation. However, for a system with a singular Lagrangian (a constrained system) a general covariant formulation is not well-established yet. We show in section 5 that our Lagrangian approach for constrained systems is able enough to be generalized to a covariant investigation.

In section 5 the Polyakov string, General Relativity and Yang- Mills theories are studied in covariant Lagrangian approach. In particular we want to emphasize on the novelty of our Lagrangian analysis of the Polyakov string. As we will see, our Lagrangian approach is sometimes much more easier and transparent in comparison with the standard Dirac approach in Hamiltonian formalism.

\section{FPS decomposition} 
 Consider a dynamical system with $P$ degrees of freedom described by the Lagrangian $L(q, \dot{q}, t)$. The Euler-Lagrange equations of motion read $L_{i}=0$ where $L_{i}$'s, denoted as Eulerian driavetives, are as follows
 \begin{equation}
 L_{i}\equiv \dfrac{d}{dt}(\dfrac{\partial L}{\partial \dot{q}_{i}})-\dfrac{\partial L}{\partial q_{i}}, \qquad\qquad{i=1,......,P}.\label{03}
 \end{equation}
 Using the Hessian matrix, defined as
\begin{equation}
W_{ij}=\dfrac{\partial^{2 L}}{\partial\dot{q}_{i}\partial\dot{q}_{j}} , \label{04}
\end{equation}
the Eulerian derivatives can be written as
\begin{equation}
L_{i}= W_{ij}\ddot{q}_{j}+\alpha_{i},\label{.04}
\end{equation}
where
\begin{equation}
\alpha_{i}=\dfrac{\partial^{2}L}{\partial{q}_{j}\partial\dot{q}_{i}}\dot{q}_{j}-\dfrac{\partial L}{\partial q_{i}}.\label{05}
\end{equation}

For ordinary (non singular) systems the Hessian matrix can be inverted to give the
 accelerations in terms of the coordinates and velocities. However, if $\det W=0$, the Lgrangian is said to be \textit{singular}; this prevents the whole set of accelerations to be determined in terms of coordinates and velocities. Suppose the rank of $W$ is $(P-A_0)$, leading to $A_0$ null-vectors $\lambda^{a_{0}}$ such that
\begin{equation}
\lambda^{a_{0}}_{i} W_{ij}=0, \hspace{1.2 cm}
a_0=1,\cdots A_0.\label{06}
\end{equation}
 Multiplying both saids of Eq. (\ref{.04}) by $\lambda_{i}^{a_0}$ gives the following equations
\begin{equation}
\Gamma^{a_{0}}(q, \dot{q})= \lambda^{a_0}_{i}L_{i} = \lambda^{a_{0}}_{i}\alpha_{i}\approx0, \label{07}
\end{equation} 
where in the last step the symbol $\approx$ means weak equality, i.e.  equality on the  constraint surface.  
Assume for combinations $\lambda^{f_{0}}$ of $\lambda^{a_0}$'s we have identities $\lambda^{f_0}_{i} \alpha_{i}= \lambda^{f_0}_{i}L_{i}=0$, where $f_0=1,\cdots F_0$. These identities are the leading set of Noether identities. The reminding equations of (\ref{07}) corresponding to null vectors $\lambda^{p_0}$ give the so called primary or first level Lagrangian constraints denoted by $\Gamma^{p_0}$'s, where $p_0=1, \cdots , P_0$. Hence, at this level we have $A_0=F_0+P_0$. By a Lagrangian constraint we mean a  function of coordinates and velocities which vanishes due to equations of motion. In other words, it is not imposed by hand from outside; instead, it emerges naturally from the dynamics of the theory.

The primary Lagrangian constraints $\Gamma^{p_{0}}$'s should remain valid during the time. Hence, the equations $d\Gamma^{p_{0}}/dt=0$ should be considered together with the original equations of motion. Assuming the constraints $\Gamma^{p_0}$ do depend on the velocities, the added equations would depend linearly on the accelerations. If some constraints depend only on the coordinates, we should consider their second time derivatives, instead. We will come back to this point later. So, the whole set of equations of motion can be written as 
\begin{equation}
W_{i_{1}j}^{1}\ddot{q_{j}}+\alpha_{i_{1}}=0,\qquad\qquad i_{1}=1,\cdots , P, P+1, \cdots , P+P_{0},
\end{equation}
where the first $P$ lines of the rectangular matrix $W^1$ is the same as the matrix $W$ and the subsequent lines from $P+1$ to $P+ P_{0}$ are in fact $\dot{\Gamma}^{p_{0}}=0$ for $p_0=1, \cdots , P_0$. 

The extended Hessian matrix $W^{1}$ may have "new null-vectors". By new null-vectors we mean null-vectors with some non vanishing element in the first $P$ components as well as the subsequent $P_{0}$ components. Consider $\lambda^{a_{1}}_{i_{1}}$, for $a_1=1,\cdots A_1$, as the components of the new null-vectors $\lambda^{a_{1}}$ of  {$\mathrm{W}^{1}$}.  In general $A_1 \le P_0$; hence the rank of (extended) Hessian matrix would be increased  by $S_1=P_0-A_1$ due to added equations of  consistency of primary constraints. Similar to the first step, the null vectors $\lambda^{a_1}$ would be separated into $F_1$ null vectors $\lambda^{f_{1}}$ such that $\lambda_{i_{1}}^{f_{1}}L_{i_{1}} \approx0$ (identically) and $P_1$ null vectors $\lambda^{p_{1}}$ where $\Gamma^{p_{1}}\equiv\lambda_{i_{1}}^{p_{1}}L_{i_{1}}$ are the second level Lagrangian constraints. 
The new Noether identities $\lambda_{i_{1}}^{f_{1}}L_{i_{1}}\approx0$ can be written in the form
\begin{equation}
\lambda^{f_1}_i L_i+\sum_{p_0=P+1}^{P+P_0}  \lambda^{f_1}_{p_0}\dfrac{d}{dt}(\lambda_j^{p_{0}} L_j)=0, \hspace{15mm} f_1=1, \cdots F_1.\label{n.neo.idn}
\end{equation} 
For future use, it is possible to find some functions $\rho_{2i}$ and $\rho_{1i}$ (see Ref. \cite{dr} for detailes) such that
 \begin{equation}  
 \dfrac{d}{dt}(\rho_{2i}L_{i})+\rho_{1i}L_{i}=0. \label{neoth}
  \end{equation} 
  Then we should investigate consistency of the  constraints $\Gamma^{p_{1}}$'s, and repeat the same procedure. 

Following an inductive approach, let us see what happens at a typical level $n$. Suppose we have obtained $P_{n-1}$ constraints $\Gamma^{p_{n-1}}$ so far, i.e.  at the $n$th level of consistency. By adding the new equations  $d\Gamma^{p_{n-1}}/dt=0$ to the previous ones, the extended Hessian matrix $W^{n-1}$ improves to $W^n$.  The dynamical equations of the system then read
\begin{equation}
W^{n}_{i_{n}j}\ddot{q_{j}}+\alpha_{i_{n}}=0,\label{08}
\end{equation}
where the row index $i_n$ of $W^n$ runs over $P+P_0+ \cdots P_{n-1}$ items corresponding to $P$ original equations of motion, $P_0$ indices $p_0$, $P_1$ indices $p_1, \cdots ,$ and $P_{n-1}$ indices $p_{n-1}$. 

The rank of the extended Hessian matrix may be increased by $S_{n}=P_{n-1}-A_{n}$ where $A_{n}$ is the number of the new null vectors $\lambda^{a_{n}}$ of the extended Hessian matrix. The new null-vector $\lambda^{a_{n}}$ should necessarily include nonzero components among the last $P_{n-1}$ indices which correspond to added lines due to $d \Gamma^{p_{n-1}}/dt=0$ as well as the first $P$ indices corresponding to the original equations of motion. However, it is possible to make other components of the null-vectors $\lambda^{a_n}$ (except the first $P$ components) vanish. This is because we are allowed to combine the previous null vectors (with enough zeros added at their tails) with a given new null-vector.

As before, the null-vectors $\lambda^{a_{n}}$  may be  divided                                                                                                                                                                                                                                                                                                                                                                                                                                                                                                                                                                                                                                                                                                                                     into $F_{n}$ null vectors $\lambda^{f_{n}}$ such that $\lambda^{f_{n}}_{i_{n}}\alpha_{i_{n}} =\lambda_{i_{n}}^{f_{n}}L_{i_{n}}\approx0$ and $P_{n}$ null vectors $\lambda^{p_{n}}$ where $\Gamma^{p_{n}}\equiv\lambda_{i_{n}}^{p_{n}}L_{i_{n}}$ are the $(n+1)$th level Lagrangian constraints.  In this way the constraints $\Gamma^{p_{n-1}}$ of the level $n$ are classified temporally into three categories as follows:

i) The F-type constraints $\Gamma^{f_n}$ , which we denote them as {\it first class Lagrangian constraints},                  corresponding to the F-type null vectors $\lambda^{f_{n}}$  which lead (upon consistency) to  Noether identities 
 \begin{equation}
 \sum^{n}_{s=0}\dfrac{d^{s}}{dt^{s}}(\rho_{si}L_{i})=0.\label{09}
 \end{equation}
 As we will see, the first class Lagrangian constraints generate  the guage symmetries of the system. 

ii) The S-type constraints $\Gamma^{s_{n}}$, which we denote them as {\it second class Lagrangian constraints}, where $d \Gamma^{s_{n}} /dt$ correspond to new independent equations for determining the accelerations. 

iii) The P-type (pending) constraints $\Gamma^{p_n}$ , corresponding to P-type null vectors $\lambda^{p_{n}}$ which lead to the next level constraints $\Gamma^{p_n}$. Note that for first and second class constraints $\Gamma^{f_n}$ and $\Gamma^{s_{n}}$ we have no subsequent constraints.

As the result of the above {\it FPS decomposition} we have 
 \begin{equation}  
P_{n-1} = F_{n} +P_{n} + S_{n}=A_n+ S_n. \label{gks}
 \end{equation}
Fig.1 is a schematic explanation to visualize what happens. Note that at the zeroth level there is no constraint; instead, we have two types of null vectors for $W$ labeled by $p_0$ and $f_0$ superscripts respectively. Hence, we have $S_0=0$ and $ A_0=P_0+F_0$. The constraints begin from the first level where $P_0$ first level constraints divide into $F_1$ first class, $S_1$ second class and $P_1$ pending constraints responsible to produce $P_1$ second level constraints; and so on.

 Now the question is what is the physical role of the pending constraints at a given level? Do they contribute to the guage symmetries or do they act as second class constraints which increase the rank of the Hessian matrix? The answer depends on what happens to the descendants of these constraints in the subsequent levels. In fact, the pending constraints do not remain pending forever. At each level of consistency a number of them would be converted to first class and a number to second class.
 
 To see what happens, consider the pending constraints $\Gamma^{p_n}$ at the $n$th level. Each combination of pending constraints would be a pending constraint. Assume the combination 
  $${\tilde{\Gamma}}^{p_{n}} \equiv \sum_{p'_{n}=1}^{P_n}  N^{p_{n}}_{p'_n} \Gamma^{p'_n}.$$ 
Under consistency process we have 
  $$\dfrac{d}{dt} {\tilde{\Gamma}}^{p_{n}}   \approx  \sum_{p'_n=1}^{P_n}  N^{p_{n}}_{p'_n} \dfrac{d}{dt} \Gamma^{p'_n} . $$
where the weak equality "$\approx$" means equality on the constraint surface. This simple calculation shows that the operations "consistency" and "combination" do commute.  Now remember from the previous page that the $(n+1)$th level constraints may emerge as a combination of the original equations of motion and the last set of $P_n$ consistency equations as follows
 \begin{equation}
 \Gamma^{p_{n+1}}=\lambda^{p_{n+1}}_{i_n} L_{i_n}=\lambda^{p_{n+1}}_{i} L_{i} +\lambda^{p_{n+1}}_{p_{n}}  d\Gamma^{p_n}/dt. \label{Gamma-n}
 \end{equation}
 Consider the redefined $n$th level constraint 
 $${\tilde{\Gamma}}^{p_{n}} \equiv \sum_{p_{n}=1}^{P_n}  \lambda^{p_{n+1}}_{p_n} \Gamma^{p_n}.$$ 
 Hence, we have 
 \begin{equation}  
\Gamma^{p_{n+1}}=\dfrac{d}{dt}  {\tilde{\Gamma}}^{p_{n}} + \lambda^{p_{n}}_{i} L_{i} . \label{desc}
 \end{equation}
This means that the $(n+1)$th constraint $\Gamma^{p_{n+1}}$ is the daughter of the $n$th level constraint ${\tilde{\Gamma}}^{p_{n}} $ which is itself the daughter of the $(n-1)$th constraint $\tilde{{\tilde \Gamma}}^{p_{n-1}} $, and so on.  This important result shows that it is, in principal,  possible to construct a chain structure in the constraints of a system as 
 \begin{equation}  
 \cdots \leftarrow \Gamma^{p_{n+1}} \leftarrow {\tilde \Gamma}^{p_n} \leftarrow \tilde{{\tilde \Gamma}}^{p_{n-1}} \leftarrow \cdots \label{sequ}
 \end{equation}
where the symbol $\leftarrow$ means "is resulted under consistency condition from". Notice that the equation (\ref{desc}) enables us, in fact, to go backward in the process of consistency of constraints as indicated by the sequence (\ref{sequ}). Each sequence or chain of constraints indicated in Eq. (\ref{sequ}) is one of the vertical columns of Fig. 1. 
\begin{center}
	\includegraphics[scale=0.6]{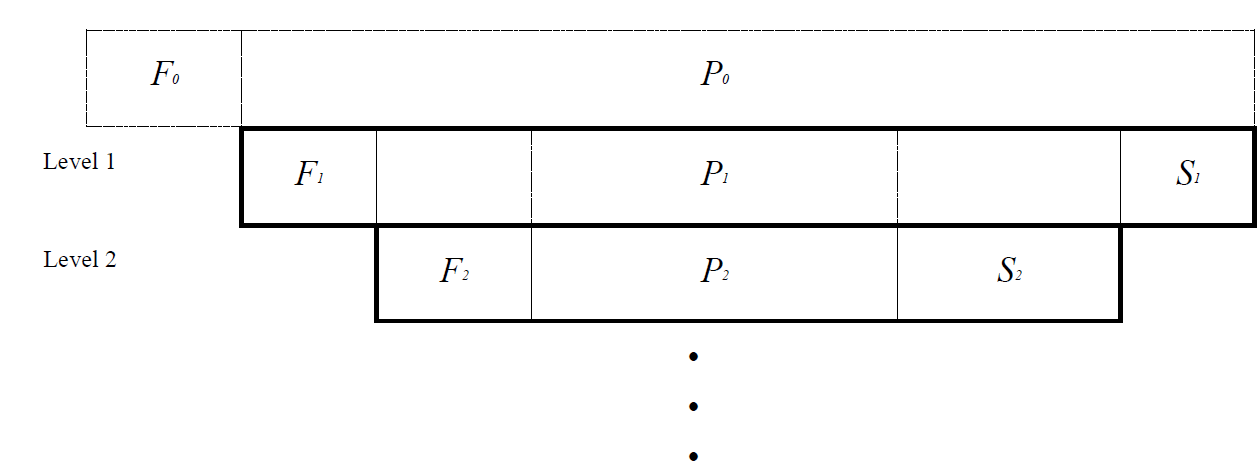}

Fig. 1 -  Schematic diagram of FPS decomposition of Lagrangian constraints
\end{center}
\vspace{3mm}

Now let us concentrate  again on the FPS decomposition of constraints at the $n$th level. Assume that, in principal, we can recombine the constraints $\Gamma^{p_n}$ such that the set $\Gamma^{n,s_n} = \sum_{p_n=1}^{P_n} \lambda^{s_n}_{p_n} \Gamma^{p_n}$ are the $n$th level second class constraints. This means that $(d/dt ) \Gamma^{n,s_n}=0 $ are new independent
equations with respect to the accelerations which increase the rank of the extended Hessian matrix by $S_n$. Following the procedure showed in Eq. (\ref{sequ}) and slightly changing the notation, we can construct the following second class chains (each with $n$ elements) 
 \begin{equation}  
 \Gamma^{n,s_{n}} \leftarrow  \Gamma^{n-1,s_{n}} \cdots \leftarrow \Gamma^{1, s_n}  \hspace{1cm} s_n=1 \cdots n. \label{ssequ}
 \end{equation}
This means that as soon as we find the set of $S_n$ second class constraints $\Gamma^{n,s_{n}} $ among the pending constraints $\Gamma^{p_n}$, we should go back to the previous level and find their parents $\Gamma^{n-1,s_{n}}$ among $\Gamma^{p_{n-1}}$; then we should go back one level further and find their ground parents $\Gamma^{n-2,s_{n}}$ among the pending constraints $\Gamma^{p_{n-2}}$, and so on to the first level. In this way we have constructed $S_n$ second class constraint chains each containing $n$ elements.

In the same way, consider the constraints $\Gamma^{n,f_n} = \sum_{p_n=1}^{P_n} \lambda^{f_n}_{p_n} \Gamma^{p_n}$ as the $n$th level first class constraints. This means that $d \Gamma^{n,f_n}/dt$ is a linear combination of the equations of motion. Using the recipe of Eq. (\ref{sequ}) we can construct similarly the following first class constraint chains
\begin{equation}  
	\Gamma^{n,f_{n}} \leftarrow  \Gamma^{n-1,f_{n}} \cdots \leftarrow \Gamma^{1, f_n}  \hspace{1cm} f_n=1 \cdots n. \label{ffequ}
\end{equation}

It is also important to note that a linear combination of each above  types of constraints remains in the same type. Hence, whenever we encounter a first class or a second class constraint at some level of consistency, we can lift vertically along its history  and indicate the corresponding ( first or second class) parents. In this way when a chain terminates by introducing a Neother identity not only the last element, but also the whole set of constraints of the chain are first class. In the same way, when a chain terminates by introducing an independent equation for determining accelerations, then all of the constraints of the corresponding chain are second class.

Hence, at the final step, where there is no pending constraint, the schematic table of constraints resembles Fig. 2 where all of the constraints are either first class or second class. The first class chains are located at the left hand side and the second class chains are located at the righty hand side of the graph. This graph corresponds to a specific case where at the final level we have both first class and second class constraints. Of course, it is possible that the longest first class chains have $N_1$ elements and the longest second class chains have $N_2$ elements and $N_1 \ne N_2$.
\begin{center}
	\includegraphics[scale=1]{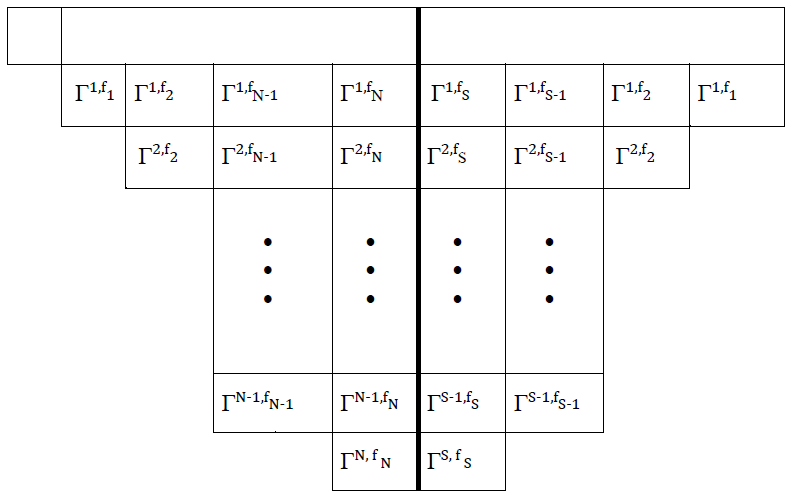}
	Fig.2 - Final chain structure of the Lagrangian constraints
\end{center}
\section{Degrees of freedom count}
Let us first see how many degrees of freedom would be decreased due to first class Lagrangian constraints. The Neother identities derived at different leves of consistency procedure, labeled by the index $f$, may be written as 
 \begin{equation}
 \sum^{n_f}_{s=0}\dfrac{d^{s}}{dt^{s}}(\rho^{(f)}_{si}L_{i})=0,\label{neother-iden}
 \end{equation}
 where $n_{f}$ is the number of Lagrangian constraints in the corresponding chain. The index $f$ takes value among $F=F_0+F_1 + \cdots +F_{N_1}$ integers and indicates the vertical column of the first class constraints corresponding to a Neother identity  (the left hand side columns in Fig. 2). This index, on the other hand, enumerates different gauge functions of the system. As shown in \cite{dr}, given a Neother identity of the form (\ref{neother-iden}),  one can show that the action, as well as the equations of motion, are unchanged  under the following gauge transformation
\begin{equation}
\delta q_{i}=\sum_{f=1}^{F}\sum^{n_{f}}_{s=0}(-1)^{s}\dfrac{d^{s}{\eta^{f}}}{dt^{s}}\rho_{si}^{(f)},\label{0010}
\end{equation}
where the arbitrary functions $\eta^{f}(t)$ are the corresponding gauge functions. According to Eq. (\ref{0010}), due to the gauge symmetry labeled by $f$ there exist $n_f+1$ independent parameters  $\eta(t), \dot{\eta}(t), \ddot{\eta}(t), \cdots , d^n\eta(t)/dt^n$ which decrease the number of free dynamical parameters by the same amount. Adding  all gauge symmetries together, the total number of guage parameters reads
\begin{equation}
\sum_{f=1}^{F} (n_f+1)=F+F.C \label{No.FC}
\end{equation}
where $F.C=\sum_f n_f$ is the total numbers of first class Lagrangian constraints. This contribution of the first class constraints should be subtracted from the total number of the original degrees of freedom. 

Now let us see what is the contribution of second class constraints. Each second class Lagrangian constraint puts one limitation on the accessible  region of space of coordinates and velocities denoted by TQ. Roughly speaking, due to each second class constraint one coordinate or one velocity would be frozen; or in other words, one of the initial conditions is no longer arbitrary. However, in a dynamical theory with second order differential equations of motion, each dynamical degree of freedom corresponds to two initial conditions. By a dynamical variable we mean a time dependent quantity with arbitrary initial value and rate of change, where its subsequent values are determined uniquely due to equations of motion. 

Hence, the number of the dynamical degrees of freedom decreases by half of the number of second claass constraints. Our final formula for the number of dynamical degrees of freedom reads
\begin{equation}
D= P-(F.C+F)-\frac{1}{2}S.C,  \label{No.dyn}
\end{equation}
where $(F.C+F)$ is the number of guage parameters (as indicated in Eq. (\ref{No.FC})) and $S.C$ is the total number of second class constraints. This is the most important formula of this paper. It resembles to the famous formul (\ref{01}) in the Hamiltonian formulation. 

An important subtlety arises here concerning the constraints which depend only on coordinates. Such constraints give velocity dependent constraints upon differentiation with respect to time. Our analyses in the previous section was based on velocity dependent constraints. Hence, whenever we find a velocity independent constraint we should differentiate it once with respect to time and take it into account, in our constraint analysis,  as an ordinary Lagrangian constraint. The main constraint only survives as an additional constraint which subtracts one initial condition. Hence, the number $S.C$ in formula (\ref{No.dyn}) should also include the number of velocity independent constraints, say $C$. In this way the number $S.C$ in formula (\ref{No.dyn}) reads
\begin{equation}
S.C= \sum_{n=1}^{N_2} S_n + C.  \label{S-number}
\end{equation}
\section{Examples}
The first three examples of this section are simple toy examples which helps the reader to capture the main aspects of the previous sections more rapidly. The last example is the electromagnetic
\vspace{0.2 cm}
 theory in a non covariant approach.\\ \vspace{0.2 cm}
\textbf{Example 1}\\ 
Consider the Lagrangian
\begin{equation}
L=\frac{1}{2}\dot{q}^2_1+\dot{q}_2 q_1+\dot{q}_3q_2+\frac{1}{2}q^2_3.  \label{ex1-L}
\end{equation}
The equations of motion (\ref{ex1-L}) read
\begin{equation}
\left( {\begin{array}{*{20}{c}}
	{1} & {0} & {0} \\
	{0} & {0} & {0} \\
	{0} & {0} & {0} \\
	\end{array}} \right)
\left( {\begin{array}{*{20}{c}}
	{\ddot q}_1\\
	{\ddot q}_{2}\\
	{\ddot q}_{3}\\
	\end{array}}\right)+
\left( {\begin{array}{*{20}{c}}
	{-\dot{q}_{2}}\\
	{\dot {q}_{1}-\dot{q}_{3}}\\
	{\dot q}_2-q_3\\
	\end{array}}\right)=0.\label{ex2-L}
\end{equation}
Multiplying from the left  by the null-vectors $(0, 1, 0)$ and $(0, 0, 1)$, gives the first level Lagrangian constraints  $\Gamma^1={\dot q}_1-{\dot q}_3 $ and  $\Gamma^2={\dot q}_2-q_3$. Annexing the time derivatives of the constraints to Eqs (\ref{ex2-L}) gives
\begin{equation}
\left( {\begin{array}{*{20}{c}}
	{1} & {0} & {0} \\
	{0} & {0} & {0} \\
	{0} & {0} & {0} \\
	{1} & {0} & {-1}\\
	{0} & {1} & {0}\\
	\end{array}} \right)
\left( {\begin{array}{*{20}{c}}
	{\ddot q}_{1}\\
	{\ddot q}_{2}\\
	{\ddot q}_{3}\\
	\end{array}}\right)+
\left( {\begin{array}{*{20}{c}}
	{-q_{2}}\\
	{\dot {q}_{1}-\dot{q}_{3}}\\
	{\dot q}_2-q_3\\
	{0}\\
	{-\dot{q}_{3}}
	\end{array}}\right)=0.\label{ex3-L}
\end{equation}
The extended Hessian matrix has maximal rank 3. Hence the constraints $\Gamma^{1}$ and $\Gamma^{2}$ are second class. In fact, we have three independent equations to determine all accelerations. However, the initial conditions and subsequent dynamics should be consistent with constraints $\Gamma^{1}$ and $\Gamma^{2}$. The number of dynamical degrees of freedom from the master formoul (\ref{No.dyn}) reads $D=3-\dfrac{1}{2}\times 2=2$.  One can omit the variable $q_{3}$ from the very beginning, to find two independent dynamical equations $\ddot{q}_{1}-\dot{q}_{2}=0$ and $\dot{q}_1-\ddot{q}_2=0$, which is uniquely solved in term of $4$ of initial values. $q_{3}$ is then determined via the constriant $\Gamma^{2}=\dot{q}_{2}-{q_{3}}=0$.

It worth noting that the cononical Hemiltonian of the system reads $H_{c}=\dfrac{1}{2}{p_{1}}^{2}-\dfrac{1}{2}{q_{3}}^{2}$ and we have two primary constraints $\chi_{1}=p_{2}-q_{1}$ and $\chi_{2}=p_{3}-q_{2}$ which are second class. Consistency of $\chi_{1}$ and $\chi_{2}$ determines the Lagrange multiplaiers of the total Hamiltonian $H_{T}=H_{C}+u\chi_{1}+v\chi_{2}$ as $u=0$ and $v=p_{1}$. The number of Hamiltonian degrees of freedom from Eq(\ref{01}) reads $D^{H}=2\times3-2=4$, as
\vspace{0.3cm}
 expected.\\
\textbf{Example 2}

\vspace{0.5cm}
Consider the Lagrangian 
\begin{equation}
L=\dfrac{1}{2}{\dot q}_{1}^{2}+\dot{q}_{2}(q_{1}-q_{2}).
\end{equation}
Multiplying the equations of motion
\begin{equation}
\left( {\begin{array}{*{20}{c}}
	{1} & {0}  \\
	{0} & {0} \\
	\end{array}} \right)
\left( {\begin{array}{*{20}{c}}
	{\ddot q}_1\\
	{\ddot q}_{2}\\
	\end{array}}\right)+
\left( {\begin{array}{*{20}{c}}
	{-\dot{q}_{2}}\\
	{\dot {q}_{1}}\\
	\end{array}}\right)=0\label{ex2-L}
\end{equation}
 with the null-vector $(0, 1)$ gives the first level Lagrangian constraint $\Gamma^{1}={\dot{q}}_{1}$. Annexing the equation $d\Gamma^{1}/dt=0$ gives the extended Hessian  
\begin{equation}
W^{1}=\left( {\begin{array}{*{20}{c}}
	{1} & {0}\\
	{0} & {0}\\
	{1} & {0}\\
	\end{array}}\right)\nonumber
\end{equation}
which has the new null-vector $(1, 0, -1)$. Multiplying the
system by this null-vector gives the second level Lagrangian constraint $\Gamma^{2}=-\dot{q}_{2}$. Differentiating $\Gamma^{2}$ with respect to time gives the independent equation $\ddot{q}_{2}=0$ for accelerations. The extended Hessian 
\begin{equation}
\left( {\begin{array}{*{20}{c}}
	{1} & {0}\\
	{0} & {0}\\
	{1} & {0}\\
	{0} & {1}\\
	\end{array}}\right)\nonumber
\end{equation}
is now full rank. However, we have a chain of second class constriants as $\left( {\begin{array}{*{20}{c}}
	{ \dot{q}_{1}}\\
	{-\dot{q}_{2}}\\
	\end{array}}\right)$. Hence the number of degrees of freedom is $D=2-\dfrac{1}{2}\times2=1$. 

In a canonical treatment of the problem we have the primary constraint $\phi_{1}=p_{2}-q_{1}+q_{2}$ and the total Hamiltonian $H_{T}=\dfrac{1}{2}p_{1}^{2}+u(p_{2}-q_{1}+q_{2}).$ Consistency of the primary constraint gives the secondry constraint $\phi_{2}=-p_{1}.$ Hence, we have a two level second class constraint chain $\left( {\begin{array}{*{20}{c}}
	{\phi_{1}}\\
	{\phi_{2}}\\
	\end{array}}\right)$ 
\vspace{0.5cm} and the number of dynamical degrees of freedom is $D^{H}=2\times2-2=2$.\\
\textbf{Example 3}
\vspace{0.5cm}

Consider the Lagrangian 
\begin{equation}
L=\dfrac{1}{2}({{\dot{q}}^2}_1+{{\dot{q}}^2}_2)+{{\dot{q}}}_1{{\dot{q}}}_2+{{\dot{q}}}_3({\dot{q}}_1+{{\dot{q}}}_2)-\dfrac{1}{2}q^{2}_1+\dfrac{1}{2}q^{2}_2. \label{exl-4}
\end{equation}
The equations of motion read 
\begin{align}
\left( {\begin{array}{*{20}{c}}
	{1} & {1} & {1} \\
	{1} & {1} & {1} \\
	{1} & {1} & {0} \\
	\end{array}} \right)
\left( {\begin{array}{*{20}{c}}
	{\ddot q}_{1}\\
	{\ddot q}_{2}\\
	{\ddot q}_{3}\\
	\end{array}}\right)+
\left( {\begin{array}{*{20}{c}}
	{q_{1}}\\
	{-q_{2}}\\
	{0}\\
	\end{array}}\right)=0.\label{ex4-L}
\end{align}
 The Hessian matrix has null-vector $(1, -1, 0)$ in the zeroth level. Multiplying both sides of  Eq. (\ref{ex4-L}) by this null-vector, gives the first level constraint $\Gamma^{1}=q_{1}+q_{2}$, which does not contain velocities. This implies the second level constraint $\Gamma^{2}=\dot{q}_{1}+\dot{q}_{2}$ which gives $\ddot{q}_{1}+\ddot{q}_{2}=0$ upon differentiation. Hence the extended equations of motion read
\begin{align}
\left( {\begin{array}{*{20}{c}}
	{1} & {1} & {1} \\
	{1} & {1} & {1} \\
	{1} & {1} & {0} \\
	{1} & {1} & {0} \\
	\end{array}} \right)
\left( {\begin{array}{*{20}{c}}
	{\ddot q}_{1}\\
	{\ddot q}_{2}\\
	{\ddot q}_{3}\\
	\end{array}}\right)+
\left( {\begin{array}{*{20}{c}}
	{q_{1}}\\
	{-q_{2}}\\
	{0}\\
	{0}\\
	\end{array}}\right)=0.\label{ex4.-L}
\end{align}
The extended Hessian matrix has the new null-vector $(0, 0, -1, 1)$ which gives an identity upon multiplying by  Eq. (\ref{ex4.-L}). Remembering that the last line of Eq.  (\ref{ex4.-L}) is in fact $d^2(L_1-L_2)/dt^2$, this identity means  $L_{3}-d^2(L_1-L_2)/dt^2=0$. Comparing this equation with the Neother identity (\ref{neother-iden}) gives
\begin{equation}
\rho_{0i}=\delta_{3i},\hspace{1cm}
\rho_{2i}=\delta_{1i}+\delta_{2i}.
\end{equation}
The gauge variations of the variables  can be written directly from Eq. (\ref{0010}) as 
\begin{equation}
\delta q_{1}=-\delta q_{2}= -\ddot{\eta}(t), \hspace{1cm}
\delta q_{3}=\eta(t).
\end{equation}
In this problem we have one two-level chain of first class constraint and no second class constraint. Taking into account $F=2$ and $G=1$ we have $D=3-(2+1)=0$ as the number of dynamical degree of freedom. In fact the equations of motion are limited to $q_{1}=-q_{2}={\ddot{q}}_3$, whose solution is $q_{3}=f(t)$ and $q_{1}=-q_{2}={\ddot{f}}$ for arbitrary $f$. Hence no initial condition is needed to fix the solution. In the Hamiltonian language, we have three first class constraints $\phi_{1}=p_{1}-p_{2}$, $\phi_{2}=q_{1}-q_{2}$, $\phi_{3}=p_{3}$
\vspace{5 mm}
 and zero number of dynamical variables (via Eq. (\ref{01})).\\
{\bf Example 4 Electromagnetism} 
 \vspace{5mm}
 
Consider the well-known action of electromagnetism as
\begin{equation}
S=-\dfrac{1}{4} \int d^{4}x  F^{\mu\nu}(x) F_{\mu\nu}(x),\label{011}
\end{equation} 
where, $A^\mu(x)$ are four field variables.  The equations of motion read $L^{\mu}=0$, where $L^\mu$ are the following Eulerian derivatives 
\begin{equation}
L^{\mu} \equiv -\partial_{\nu}F^{\mu\nu} =-\partial_{\nu} (\partial^{\mu} A^{\nu} -\partial^{\nu}A^{\mu}).\label{012}
\end{equation}
 Assuming the metric of the flat space in natural units as  diagonal  $(-1,1,1,1)$ and using the matrix notation of section 2, the equations of motion would be written as 
\begin{align} 
\left( {\begin{array}{*{20}{c}}
	{0} & {0} & {0} & {0} \\
	{0} & {1} & {0} & {0} \\
	{0} & {0} & {1} & {0} \\
	{0} & {0} & {0} & {1} \\
	\end{array}} \right) \left( \begin{array}{l}
\partial_0 \partial^0A^{0}\\ 
\partial_0 \partial^0A^{1}\\ 
\partial_0 \partial^0A^{2}\\ 
\partial_0 \partial^0A^{3}\\ 
\end{array} \right)
+
\left( \begin{array}{l}
\partial_{i} \partial^{i}A^{0}-\partial_{i}\partial^0A^{i}\\ 
-\partial^{1} \partial_{0}A^{0} + \partial_{i} \partial^{i}A^{1}-\partial_{i}\partial^1A^{i}\\ 
-\partial^{2} \partial_{0}A^{0} + \partial_{i} \partial^{i}A^{2}-\partial_{i}\partial^2A^{i}\\ 
-\partial^{3} \partial_{0}A^{0} + \partial_{i} \partial^{i}A^{3}-\partial_{i}\partial^3A^{i}\\ 
\end{array} \right)=0.\label{013}
\end{align}
The null eigenvector of the Hessian in Eq.  (\ref{013}) is $\lambda^0=(1, 0, 0, 0)$ which gives the first level Lagrangian constraint as
\begin{equation}
\gamma^{1}\equiv \partial_{i}\partial^{i}A^{0} -\partial_{i}\partial_{0}A^{i}=L^0.\label{015}
\end{equation}
Adding $L_{5} \equiv \dfrac{\partial\gamma}{\partial t}=\partial_0 L^0$ to the previous equations of motion gives
\begin{align} 
\left( {\begin{array}{*{20}{c}}
	{0} & {0} & {0} & {0} \\
	{0} & {1} & {0} & {0} \\
	{0} & {0} & {1} & {0} \\
	{0} & {0} & {0} & {1} \\
	{0} & {-\partial_{1}} & {-\partial_{2}} & {-\partial{3}}\\
	\end{array}} \right) \left( \begin{array}{l}
\partial_0 \partial^0A^{0}\\ 
\partial_0 \partial^0A^{1}\\ 
\partial_0 \partial^0A^{2}\\ 
\partial_0 \partial^0A^{3}\\ 
\end{array} \right)
+
\left( \begin{array}{l}
\partial_{i} \partial^{i}A^{0}-\partial_{i}\partial^0A^{i}\\ 
-\partial^{1} \partial_{0}A^{0} + \partial_{i} \partial^{i}A^{1}-\partial_{i}\partial^1A^{i}\\ 
-\partial^{2} \partial_{0}A^{0} + \partial_{i} \partial^{i}A^{2}-\partial_{i}\partial^2A^{i}\\ 
-\partial^{3} \partial_{0}A^{0} + \partial_{i} \partial^{i}A^{3}-\partial_{i}\partial^3A^{i}\\ 
\partial_{i}\partial^{i}\partial_{0}A^{0}\\
\end{array} \right)=0.\label{014}
\end{align}
The new null eigenvector of the extended Hessian reads $\lambda^{1}= (0, \partial_{1}, \partial_{2}, \partial_{3}, 1)$.  Multiplying Eq. (\ref{014}) by $\lambda^{1}$ gives an identity. Hence, we reach to the Nother identity $L^5+\partial_i L^i=0$, which can be written as 
\begin{equation}
\frac{\partial}{\partial t} (L^0) + (\partial_i L^i)=0 \label{niEM}
\end{equation}
or
\begin{equation}
\partial_{\mu}L^{\mu} =0.\label{024}
\end{equation}
One might obviously find this Neother identity by imposing the partial derivatives $\partial_\mu$ on the Eulerian derivatives $L^\mu$ given in Eq.  (\ref{012}). We will discuss in the next section the covariant approach to classical field theories. In fact, finding a Neother identity by every reasonable method, enables us to find a gauge transformation by using the mechanism explained in section 2. This may include all possible trial and error manipulations.  However, our method (of finding the null-vectors of the extended Hessian matrix) gives a systematic approach to find all the gauge symmetries together with the corresponding Lagrangian constraints which generate them. 

Now we can read directly the Lagrangian generators $\rho_{si}^{(g)}$ of gauge transformations  from the Neother identity (\ref{niEM}). Noting that in field theory every summation over the index $i$ of the Eulerian derivatives $L_i$ includes also a spacial integration over the space variables $\text{z}$, say, we have  
\begin{align}
\rho_{10}=&\delta^3(\text{z}-\text{x}),\label{025}\\
\rho_{0i}=&-\partial_{\text{z}_i}\delta^3(\text{z}-\text{x}).\label{026}
\end{align}
Inserting the above Lagrangian gauge generators into Eq. (\ref{0010}) and performing the spacial integration over the $\text{z}$-variable, gives the following gauge transformations for the field components  
\begin{equation}
\delta A^{0}=-\partial_{0}\eta(\text{x},t), \label{027}
\end{equation}
\begin{equation}
\delta A^{i}=\partial_{i}\eta(\text{x},t),\label{028}
\end{equation}
where  $\eta(\text{x},t)$ is an arbitrary field. The transformations (\ref{027}) and (\ref{028}) can be written covariantly as  
\begin{equation}
\delta A^{\mu}= \partial^{\mu}\eta. \label{029}
\end{equation}

\section{Covariant Formalism}
In this section we want to investigate the procedure of sections 2 and 3 in a covariant approach. Assume a dynamical system described by the action 

\begin{equation}
S=\int d^4 x {\cal{L}} ({\phi^a}, \partial_\mu {\phi^a}).\label{cov1}
\end{equation}
where the index ''a'' may represent a collective set of indices including tensorial or fermionic ones. The Euler-Lagrange equations of motion read 
\begin{equation}
{{\cal L}_a} \equiv \partial_\alpha \left(\dfrac{\partial \cal L}{\partial(\partial_\alpha \phi^a)}\right)-\dfrac{\partial \cal L}{\partial \phi^a}=0,\label{cov2}
\end{equation}
where ${\cal L}_a$ is the Eulerian derivative corresponding to the field $\phi^a$. Expanding the first term in Eq. (\ref{cov2}) we have 
\begin{equation}
{\cal L}_a  = {W_{ab}}^{\alpha\beta}\partial_\alpha\partial_\beta \phi^b+ A_a\label{cov3}
\end{equation}
where 
\begin{equation}
{W_{ab}}^{\alpha\beta}=\dfrac{\partial^2 {\cal L}}{\partial(\partial_\alpha \phi^a)\partial(\partial_\beta \phi^b)},\label{cov4}
\end{equation}
\begin{equation}
A_a=\dfrac{\partial^2 \cal L}{\partial(\partial_\alpha \phi^a)\partial \phi^b}\partial_\alpha \phi^b-\dfrac{\partial \cal L}{\partial \phi^a}.\label{cov5}
\end{equation}
${W_{ab}}^{\alpha\beta}$ and $A_a$ are covariant analogues of $W_{ij}$ and $\alpha_i$ of section $(2)$, respectively.

Assume there exist a null-vector $\lambda^a(x) $ where $\lambda^a {W_{ab}}^{\alpha\beta}=0. $ If $\lambda^a L_a $ vanishes identically we have a one stage Neother identity; otherwise we have a first level Lagrangian constraint as $\Gamma(\phi, \partial \phi) \equiv \lambda^a A_a=\lambda^a L_a $. As before, we can add the space-time derivatives of the primary constraint to the existing equations of motion. The whole procedure is exactly the same as what we did for systems with finite number of degrees of freedom in sections 2 and 3. However, here we have so many indeces due to space-time derivatives instead of simple dots.  Let us consider in particular the covariant form of a Neother identity as 
\begin{equation}
\sum\limits_{s=0}^n \partial_{\mu_1} \partial_{\mu_2} \cdots \partial_{\mu_s}  \left( \rho^{ (a)\mu_{1} \mu_{2}\cdots \mu_s } {\cal L}_a \right)=0,\label{cov6}
\end{equation} 
where summation over repeated covariant indices is understood. Note that the index "$s$" in Eqs. (\ref{neother-iden}) and (\ref{0010}), which indicates the number of derivatives, in no more needed to be specified in Eq. (\ref{cov6}) and in the following. In a similar way, given in ref. \cite{dr} (where one deduces the gauge transformation (\ref{0010}) from the Neother identity (\ref{neother-iden})) one can show the Lagrangian (\ref{cov1}) is invariant under the following gauge transformation
\begin{equation}
\delta \phi^{(a)}=\sum\limits_{s=0}^n (-1)^{s} \rho^{ (a)\mu_{1} \mu_{2}\cdots \mu_s } {\partial_{\mu_{1}}\partial_{\mu_{2}}\cdots\partial_{\mu_{s}}}\eta,\label{cov7}
\end{equation}
where $\eta(\text{x},t)$ is an arbitrary field. If there are several Neother identities, enumerated by the collective  index $k$, we may identify the corresponding arbitrary fields by $\eta^k(\text{x},t)$. 

For a field theory over a curved space-time it is just needed to change ordinary derivatives $\partial_\mu $ to the covariant derivatives $\nabla_\mu $ in all equations from (\ref{cov1}) to (\ref{cov7}). Let us particularly write down the covariant form of Eqs. (\ref{cov6}) and (\ref{cov7}) as follow 
\begin{equation}
\sum\limits_{s=0}^n \nabla_{\mu_1} \nabla_{\mu_2} \cdots \nabla_{\mu_s}  \left( \rho_{k}^{ (a)\mu_{1} \mu_{2}\cdots \mu_s } {\cal L}_a \right)=0,\label{cov8}
\end{equation} 
and
\begin{equation}
\delta \phi^{(a)}=\sum\limits_{s=0}^n (-1)^{s} \rho_{k}^{ (a)\mu_{1} \mu_{2}\cdots \mu_s } {\nabla_{\mu_{1}} \nabla_{\mu_{2}} \cdots\nabla_{\mu_{s}}}\eta^{k}.\label{cov9}
\end{equation}
Note that if the Neother identities are distinguished by tensorial indices, the same indices should be attached to the arbitrary functions $\eta^k(\text{x},t)$, then the covariant derivatives should act  appropriately. To clarify the formalism, let us analyze some examples in details.
\subsection{Polyakov string}
As an extension of the relativistic point particle, the Polyakov action of a string is as follows \cite{zwiebach}
\begin{equation}
S=-\dfrac{1}{4\pi\alpha^{'}}\int d\tau d\sigma \sqrt{-g} g^{\alpha\beta}\partial_{\alpha}X^{\mu}\partial_{\beta}X^{\nu}\eta_{\mu\nu}\label{2.10},
\end{equation}
 where $\tau$ and $\sigma$ are coordinates of the string world-sheet with the metric $g_{\alpha\beta}(\tau,\sigma)$ 
and $X^{\mu}(\tau,\sigma)$ are the so-called coordinate fields 
of a flat $d$-dimensional target space. Assuming the coordinate fields and the components of the inverse  world-sheet metric as physical variables, the Polyakov string possesses altogether $d+3$ primitive degrees of freedom. Assume the index $a $ enumerates our field variables, such that $a=1 ,\cdots, d $  denote the coordinate fields $ X^{\mu}$, and $ a=d+1,d+2 , d+3$  refer to three independent components of the inverse metric $ g^{\alpha\beta}$. The Eulerian derivatives corresponding to variables $X^\mu$ and $g^{\alpha \beta}$ are denoted as $L_\mu$ and $ L_{\alpha\beta}$ respectively and are derived as
\begin{equation}
L_{\mu}=2\partial_{\alpha}(\sqrt{-g}g^{\alpha\beta}\partial_{\beta}X_{\mu}),\label{2.14}
\end{equation}
and
\begin{equation}
L_{\rho\sigma}\equiv\sqrt{-g}k_{\rho\sigma}={\sqrt{-g}}\lbrace\partial_{\rho}X^{\mu}\partial_{\sigma}X_{\mu}-\dfrac{1}{2}g_{\rho\sigma}(g^{\gamma\delta}\partial_{\gamma}X^{\mu}\partial_{\delta}X_{\mu})\rbrace. \label{2.15}
\end{equation}

As is seen, the equations of motion (\ref{2.15}) do not contain accelerations. This means that the last $3$ rows and columns of $W_{ab}$ are zero. Rewriting Eq. (\ref{2.14}) as
\begin{equation}
\frac{1}{2}L_{\mu}=(\sqrt{-g}g^{\alpha\beta})\partial_{\alpha}\partial_{\beta} X_{\mu}+\partial_{\alpha}(\sqrt{-g}g^{\alpha\beta}) \partial_{\beta}X_{\mu},\label{2.17}
\end{equation}
the covariant Hessian matrix reads
\begin{equation}
W^{\alpha\beta}= \left(  \begin{array}{cc}
 \sqrt{-g}g^{\alpha\beta}&0\\ 0&0_{3 \times 3}\\ 
\end{array} \right) .
\end{equation}
The apparent null-vectors give directly the expressions $ L_{\rho\sigma}$ as first level Lagrangian constraints. However, due to the identity  
 \begin{equation}
 g^{\rho\sigma}L_{\rho\sigma}=0, \label{2.16}
 \end{equation}
 ${L_{\rho\sigma}}$'s are not independent functions. This is the well-known fact that the energy- momentom tensor of the Polyakov string is identically traceless. Hence, we can consider the identity (\ref{2.16}) as a Neother identity, and count on two independent first level Lagrangian constraints as independent combinations of the constraints $L_{\alpha\beta}$.
 
 According to prescription of section $2$, we should add the derivatives of the constraints to the existing equations of motion. Differentiating the constraints (\ref{2.15}) gives
 \begin{align}
 \nonumber   \partial_{\lambda}(\sqrt{-g}k_{\rho\sigma})= &\sqrt{-g} \partial_{\lambda}\partial_{\alpha} X^{\mu} \left(\delta_\rho^\alpha\partial_{\sigma}X_{\mu}+ \delta^\alpha_\sigma \partial_{\rho}X_{\mu} - g_{\rho\sigma}g^{\alpha\beta} \partial_{\beta}X_{\mu}\right)
 \\&-\dfrac{1}{2}\sqrt{-g} \partial_{\lambda}(g_{\rho \sigma} g^{\alpha \beta})\partial_{\alpha} X^{\mu}\partial_{\beta} X_{\mu} + (\partial_{\lambda} \sqrt{-g})k_{\rho \sigma}.\label{2.21}
 \end{align}
 In this way we have six further equations due to different choices of $\lambda$ and $\rho\sigma$ on the l.h.s. of Eq. \ref{2.21}. Hence we have altogether a set of $d+3+6$ equations as
 \begin{align}
 & L_\mu=0 \nonumber \\ & L_{\rho \sigma} =0 \label{poleqs} \\ & \partial_\lambda L_{\rho \sigma} =0. \nonumber
 \end{align}
  Note that  the first term on the r.h.s of this equation contains accelerations. This indicates that the extended Hessian matrix includes six new nontrivial rows. However, due to identity (\ref{2.16}), equations (\ref{2.21}) should be considered as four, instead of six, independent equations containing accelerations. In fact differentiating Eq (\ref{2.16}) gives 
 \begin{equation}
 (\partial_\lambda g^{\rho \sigma})L_{\rho \sigma}+g^{\rho \sigma}\partial_\lambda L_{\rho \sigma}=0. \label{2.116}
 \end{equation}
 Eqs. (\ref{2.16}) and (\ref{2.116}) correspond respectively to the null-vectors $(0,g^{\rho \sigma}, 0)$ and $(0, \partial_\lambda g^{\rho \sigma}, g^{\rho \sigma}) $ concerning the last $(3+6)$ rows of the extended equations of motion (\ref{poleqs}).
 
 Now, we should search for new null-vectors containing non vanishing element in the first d elements. Looking carefully on the contents of the Hessian matrix shows the existence of the following two null-vectors
 \begin{equation}
 \lambda'_{(\beta)}= \left( \partial_{\beta}X^{\mu}\ ,\  0\ ,\ - 2\delta_{\beta}^{\sigma} g^{\lambda\rho} +2\delta_{\beta}^{\lambda}g^{\rho\sigma} \right).
 \label{2.23}
 \end{equation} 
 Multiplying the extended set of equations of motion (\ref{poleqs}), from the left, by these null-vectors gives a combination of equations of motion, i. e. no further constraint, as follows
  \begin{align}
    (\partial_{\beta}X^{\mu})L_{\mu} + \left( -2\delta_{\beta}^{\sigma} g^{\lambda\rho}   +2\delta_{\beta}^{\lambda}g^{\rho\sigma} \right)  \partial_{\lambda} L_{\rho \sigma} = \left( -\partial_\beta g^{\rho \sigma} + 2\delta^\sigma_\beta \partial_\lambda g^{\lambda \rho} \right) L_{\rho \sigma}  \approx 0.\label{prenoder}
  \end{align}
 However, Eq. (\ref{prenoder}) leads to following Neother identities labeled by the index $\beta$
  \begin{equation}
  (\partial_{\beta}X^{\mu})L_{\mu}-(\partial_{\beta}g^{\rho\sigma})L_{\rho\sigma}-\partial_{\lambda}(2\delta_{\beta}^{\sigma}g^{\lambda\rho}L_{\rho\sigma})=0.\label{2.27.0}
  \end{equation}
 In this way the constraint analysis comes to its end by two first class Lagrangian constraints which are two independent combinations of $L_{\alpha\beta} $. Moreover, we have three Neother identities, i.e. Eqs.  (\ref{2.16}) and (\ref{2.27.0}), which correspond to Weyl and reparametrization of the system respectively. Remembering our master formula (\ref{No.dyn}), we have $P=d+3$, $F.C=2$ and $F=3$, giving $D=d-2$ dynamical variables \cite{pons}.

Now let us proceed to indicate the gauge symmetries of the system by considering the Neother identities. Comparing  Eqs. (\ref{2.16}) and (\ref{2.27.0}) with the standard form (\ref{cov6}) of the Neother identities, gives the corresponding Lagrangian generators of the gauge transformations as 
\begin{align}
&\rho_W^{(\rho\sigma)}=g^{\rho\sigma},\\
&\rho_{R \beta}^{(\mu)}=\partial_{\beta}X^{\mu} \label{2.28}\\
&\rho_{R\beta}^{(\rho\sigma)}=-\partial_{\beta}g^{\rho \sigma}\label{2.29}\\
&\rho_{R\beta}^{(\rho\sigma) \alpha}=-\delta_{\lambda}^{\alpha}\delta_{\beta}^{\sigma}g^{\lambda\rho}-
\delta_{\lambda}^{\alpha}\delta_{\beta}^{\rho}g^{\sigma\lambda}. \label{2.30}
\end{align}
 where the symbols $W$ and $R$ represent Weyl and reparametrization guage symmetries respectively and the indices in the parentheses is the same as index $a$ which represent the corresponding physical variable. The unwritten generators (such as $\rho_0^{W(\mu)}$) are zero. Note also there are two independent guage parameters for reparametrizations which are labeled by the index $\beta$.
 Then using the prescription (\ref{cov7}) gives the gauge transformations of physical variables as follows
 \begin{align}
 &\delta_W g^{\rho\sigma}=\Omega g^{\rho\sigma},\label{2.33.}\\
 & \delta_W X^\mu=0.\label{2.34}
 \end{align}
 and
 \begin{align}
 \delta_{R} X^{\mu}=&\eta^{\beta}\partial_{\beta} X^{\mu},\label{2.31} \\ \delta_R g^{\rho\sigma}= &-\eta^{\beta}\partial_{\beta} g^{\rho\sigma}-(-\delta_{\lambda}^{\alpha} \delta_{\beta}^{\sigma}g^{\lambda\rho}-\delta_{\lambda}^{\alpha}\delta_{\beta}^{\rho}g^{\sigma\lambda})\partial_{\alpha}\eta^{\beta} \nonumber \\
 =&-\eta^{\lambda}\partial_{\lambda}g^{\rho\sigma}+g^{\rho\lambda}\partial_{\lambda}\eta^{\sigma}+
 g^{\lambda\sigma}\partial_{\lambda}\eta^{\rho}\label{2.32},
 \end{align}
 where $\Omega$ and $\eta^\beta$ are three arbitrary fields over the world-sheet which act as guage parameters. Eqs. (\ref{2.33.}-\ref{2.32}) are the standard infinitesimal forms of the Weyl and reparametrization guage transformations respectively.
 
 \subsection{General Relativity}
 Consider the famous Hilbert-Einstein action of general relativity in a $d$-dimensional Minkowski space-time as
 \begin{equation}
 S=\int d^{d}x \sqrt{-g} R, \label{gr1}
 \end{equation}
 where $g$ is the determinant of the metric and $R$ is the scalar curvature. As is well-known, the Einstein equation of motion reads
 \begin{equation}
 G_{\mu\nu}\equiv R_{\mu\nu}-\dfrac{1}{2}Rg_{\mu\nu}=0.\label{gr2}
 \end{equation}
 It is also well-known \cite{caroll}, that the Einstein tensor $G_{\mu\nu}$ satisfies the Bianchi identity
 \begin{equation}
 \nabla^{\mu}G_{\mu\nu}=0.\label{gr3}
 \end{equation}
 In most of the text-books and papers on the dynamical content of the Einstein equation it is argued that in four dimensions we should subtract $4$ degrees of freedom due to diffeomorphism invariance of the Hilbert-Einstein action and $4$ more ones due to Bianchi identity. However, this explanation is not accurate, since these two issues are inter-connected to each other. In other words, the Bianchi identity acts as the conservation law corresponding to the reparametrization symmetry. 
 
 Fortunately the precise Hamiltonian analysis of the system, using the ADM variables shows clearly the existence of $8$ first class constraints in two levels. Accordingly  the number of dynamical degrees of freedom in phase space turns out to be $$\dfrac{1}{2}(20-2\times 8)=2.$$ Nevertheless, we think a precise Lagrangian description of the problem, needs to consider carefully the interrelation between the constraint structure and the symmetry properties of the system.
 
 In our case of general relativity the number of primitive degrees of freedom is equal to the number of independent components of the metric, i.e. $k=d(d+1)/2 $. The number of gauge symmetries is the same as the number of Bianchi identities, i.e. $G=d$, which is the same as independent gauge parameters $\varepsilon^{\mu}$ in the diffeomorphism $x^{\mu} \longrightarrow x^{\mu}+\varepsilon^{\mu}(x)$. Rewriting  the Bianichi identity (\ref{gr3}) in the form $\nabla_\rho \left[ \delta_{\mu}^{\rho} g^{\nu\lambda}{G_{\lambda}}^{\mu}+\delta_{\nu}^{\rho} g^{\mu\lambda}{G_{\lambda}}^{\nu} \right]=0$ and comparing it with the covariant  Eq. (\ref{cov8}) gives 
\begin{equation}
\rho_{1\alpha}^{(\mu \nu)\lambda}= \delta_{\alpha}^{\mu} g^{\nu\lambda}+\delta_{\alpha}^{\nu} g^{\mu\lambda},\label{gr4}
\end{equation}
 where the index "a" which enumerates the field variables (in Eqs. \ref{cov1} onward), is here the symmetric settings of $( \mu\nu)$ and the index $\lambda$ contracts with the covariant derivative (here we have only the $s=1$ term of Eq. \ref{cov8}). The free index $\alpha$ enumerates the Neother identities. Inserting ${\rho_{1(\mu \nu)\alpha}}^{\lambda}$ from Eq. (\ref{gr4}) in Eq. (\ref{cov9}) gives the variations of $g_{\mu\nu}$ as
 \begin{equation}
 \delta g_{\mu\nu}= \nabla_\mu \varepsilon_\nu+\nabla_\nu \varepsilon_\mu,\label{gr5}
 \end{equation}
 which is the well-known result for variation of metric under diffeomophism transformation.
 
 Now let us see what is the number of dynamical degrees of freedom. In particular, we need to know about the type and number of constraints. It can be seen directly that the components $R_{0\mu}$ of Ricci tensor do not include accelerations (i.e. second order time derivatives of the metric components). Hence, the combinations $R_{\mu\nu}=G_{\mu\nu}+(g^{\alpha\beta} G_{\alpha\beta})g_{\mu\nu}/2$ of Eulerian derivatives $G_{\mu\nu}$ include four acceleration-free equations. In this way there emerge, in fact, four Lagrangian constraints, which should necessarily be first class. This is so since by taking time derivatives of the constraints we should find the Bianchi identity (i.e. $\nabla_{0}G^{0\mu}+\nabla_{i}G^{i\mu}=0$) as the Neother identity of the system. So, using the master equation ({\ref{No.dyn}}) with $k=10$ and $F.C=F=4$, we find finaly $D=2$, as expected.
 
\subsection{Yang-Mills theory} 
As an important example of covariant approach to Lagrangian investigation of constrained systems, consider the Yang-Mills action given by 
\begin{equation}
S=-\dfrac{1}{4}\int d^{4}x F^{A\mu\nu}(x) F_{\mu\nu}^{A}(x),\label{3.21}
\end{equation}
where the index "$A$" runs over 1 to $|G|$, the dimension of the algebra of a given gauge group. The field strength tensor  $F_{\mu \nu}^{A}$ is written in terms of the potentials $A_{\mu}^{A}$ as \cite{peskin}
\begin{equation}
F_{\mu\nu}^{A}=\partial_{\mu}A_{\nu}^{A}-\partial_{\nu}A_{\mu}^{A}+gf^{ABC}A_{\mu}^{B}A_{\nu}^{C}.\label{3.22}
\end{equation}
The antisymmetric coefficients $f^{ABC}$ are structure constants of the given Lie algebra, as
 \begin{equation}
[T^{A}, T^{B}]= if^{ABC}T^{C}, \label{3.23}
\end{equation}
where $T^{A}$'s are generators of the corresponding gauge group. Defining the covariant derivative 
\begin{equation}
D_{\mu}= \partial_{\mu}-igA_{\mu}^{A}T^{A},\label{3.24}
\end{equation}
it is easy to see 
\begin{equation}
[ D_{\mu}, D_{\nu}]=-igF_{\mu\nu}^{A}T^{A}.\label{3.25}
\end{equation}
Varying the action (\ref{3.21}) with respect to $A^A_\mu$ gives the Eulerian derivatives as
\begin{equation}
\L^{A\mu}= -(\partial_{\nu}F^{A\mu\nu}+ gf^{ABC}A^{B}_{\nu}F^{C\mu\nu})= -(D_{\alpha}F^{\alpha \mu})^{A}.\label{3.27}
\end{equation}
Similar to general relativity, we can show directly\cite{sundermeyer}
\begin{equation}
\Omega^{A}=-D_{\mu} \L^{A\mu }=0.\label{3.28}
\end{equation}
This is a set of $|G|$ Noether identities. Rewriting Eq. (\ref{3.28}) in the detailed form 
\begin{equation}
\Omega^{A}=-\partial_{\mu}L^{A\mu }-gf^{ABC} A_{\mu}^{B} L^{C\mu},\label{3.29}
\end{equation}
and comparing it with Eq. (\ref{cov6}), we get
\begin{equation}
\rho^{A( C}_{\ \ \ \ \mu )}=gf^{ABC}A^B_\mu \hspace{1cm}
\rho^{A( C\ \ \nu}_{\ \ \ \ \mu )}= \delta^{AC}  \delta^\nu_\mu \label{3.31}
\end{equation}
as $s=0$ and $s=1$ contributions to Eq. (\ref{cov6}). Inserting the guage generators (\ref{3.31}) in Eq. (\ref{cov7}) gives the following gauge transformation
\begin{equation}
\delta A_{\mu}^C= gf^{ABC}A^B_\mu \eta^A -\partial_\mu \eta^C\label{3.32}
\end{equation}
where $\eta^C$ are gauge parameters. 

Let us find the number of degrees of freedom. The number of primitive degrees of freedom is $k=|G|d$ where d is the dimension of space-time. The number of gauge parametrs $\eta^{A}$ is simply $|G|$. We have also $|G|$ Lagrangian constraints emerging due to singularity of Hessian. It is easily seen from Eq. (\ref{3.27}) that the Eulerian derivativs $L_{0}^{a}$ do not include accelerations, i.e.
\begin{equation}
L^{A}_{0}=-\partial^{i}F^{a}_{0i}+gf^{abc}A^{ib}F^{C}_{i0},\label{3.33}
\end{equation}
where $F_{0i}^{A}$ and $F_{i0}^{A}$ include at most one time derivative of the fields $A^{A}_{\mu}$.

Consistency of the constraints $L_{0}^{A}$ gives the Neother identities (\ref{3.29}) under combining $\partial^{0}L_{0}^{a}$ with suitable combination of the original Eulerian derivatives. Hence, the constraints (\ref{3.33}) are first class. In this way we have $|G|$ first class and no second class Lagrangian constraints, i.e. $F=|G|$ and $S=0$. Putting all of these results in the master formula (\ref{No.dyn}) gives 
\begin{equation}
D=d|G|-(|G|+|G|)=(d-2)|G|.\label{3.34}
\end{equation}
For the abelian case of electrodynamics $|G|=1$ and $D=d-2$. So the above covariant approach of the non-abelian gauge theories survives simply to the abelian case of electrodynamics.

\section{conclusions}

In this paper we proposed a complete and detailed program for analyzing a constrained system in the framework of Lagrangian formalism. As we see, there is no Lagrangian counterpart associated to primary Hamiltonian constraints. However, subsequent levels of Hamiltonian constraints have their own projection on the Lagrangian system of constraints. We showed that first class Hamiltonian constraints have their counterparts in the Lagrangian formalism as constraints leading to Neother identities. While, second class Hamiltonian constraints correspond to Lagrangian constraints which freeze up a number of degrees of freedom. In this way we introduced for the first time the notion of first and second class Lagrangian constraints.

We also showed that it is possible to construct a chain structure in our Lagrangian analysis which resembles a similar approach in Hamiltonian investigation \cite{loranshirzad}. The main strategy is to extend the Hessian matrix by including consistency conditions of the Lagrangian constraints and then try to find its new null-vectors. At each level of consistency we have three types of constraints: first class, second class and pending constraint. The time derivatives of pending constraints give the next level constraints. However, after terminating all constraint chains, i.e. when the constraint analysis goes to its end, there is no pending constraint and the whole system would be divided into first and second class constraints. As an important achievement of our Lagrangian approach we deduced a master formula for calculating the number of Lagrangian dynamical degrees of freedom (see Eq. (\ref{No.dyn})). 

As is seen, our investigation shows that one can find the whole dynamical characteristics of a theory with no need to lift it to Hamiltonian formalism. In fact, within the Lagrangian formulation we are able in a simple and consistent way to describe the gauge symmetry, as well as non-gauge constraints of the system.

Finally, one impressive advantage of our Lagrangian approach is the ability to improve it towrds a covariant approach by upgrading the time derivative $d/dt$ to the space-time derivative $\partial_{\mu}$ (in flat space-time) or covariant derivative $\nabla_{\mu}$ (in curved space-time). This possibility is not available in Hamiltonian formalism. In addition to some toy examples in systems with finite number of degrees of freedom, we showed that our analysis exactly works for Yang- Mills theory, general relativity and Polyakov string. Although these theories and their dynamical characteristics are familiar to community, however, the method of investigating their features is new and noticeable. We think that the method given here can be used as an alternative to the well- known Dirac method for studing new complicated gauge theories in different branches of physics.


\begin{thebibliography}{99}
	
\bibitem{dirac} Dirac. P. A. M, "Lectures on Quantum Mechanics" New York, Yashiva University Press (1964).
\bibitem{sundermeyer} Sundermeyer. K, "Lecture Note in Physics" Springer- Verlag Berlin Heidelberg (1982).
\bibitem{henn} Henneaux. M, Teitelbiom. C, "Quantization of Gauge systems", Princeton University Press. Princeton, NJ (1991).

\bibitem{Batlle} Batlle. C, Gomis. J, Pons. J. M and Roman-Roy. N, "Equivalence between the Lagrangian and Hamiltonian formalism for constrained systems"  J. Math. Phys.27 2953 (1986).
\bibitem{gomhenpons}Gomis J, Henneaux M and Pons J M 1990 Class. Quantum Grav. 7 1089. 

\bibitem{gitman} D.M. Gitman and I.V. Tyutin, Quantization of fields with constraints,
Springer-Verlag, Heidelberg, 1990.


\bibitem{chaichian} Chaichian. M, Martinez D. L,"On the Noether identities for a class of systems with singular Lagrangians", J. Math. Phys. 35(12) (1994).
\bibitem{sudarshan}Sudarshan E C G and Mukunda N 1974 Classical Dynamics, A Modern Perspective (New York: Wiley- Interscience).
\bibitem{dr} Shirzad. A, "Gauge symmetry in Lagrangian formulation and Schwinger Models", J. Phys. A 31 2747-2760(1998).
\bibitem{banerjee} Banerjee. R, Rothe. H. J, Rothe. K. D,"Master equation for Lagrangian gauge symmetries", Phys. Lett. B 479 429-434 (2000).


\bibitem{ba roy sa} R. Banerjee, Debraj Roy, and S. Samanta, 
"Lagrangian generators of thePoincare gauge symmetries". Phys.Rev.,D82: (2010), p. 044012.
\bibitem{bansam} R. Banerjeeand S. Samanta, "Gauge generators, transformations and identities on a noncommutative space". Eur.Phys.J.,C51: (2007), pp. 207–215.

 \bibitem{loranshirzad} F. Loran and A. Shirzad, Int. J. Mod. Phys. A 17, 625  2002 

\bibitem{samanta} Samanta. S, "Diffeomorphism Symmetry in the Lagrangian formulation of Gravity" Int J Phys 48: 1436-1448(2009), arxive: 0708.3300.

\bibitem{zwiebach} Zwiebach. B, "A First Course in String Theory" Cambridge University Press (2004).
\bibitem{pons} Batlle. C, Gomis. J, and Pons. J. M, "Hamiltonian and Lagrangian constraints of the bosonic string" Phys. Rev. D 34, 2430. (1986)
\bibitem{caroll} Caroll. S. M, "An Introduction to General Relativity: Space time and Geometry" Addison Wesley (2003).
\bibitem{peskin} Peskin. M. E, and Schroeder. D. V, "An Introduction to Quantum Field Theory" Westview Press (1995).

\end{thebibliography}
\end{document}